\def \beq {\begin{equation}}
\def \eeq {\end{equation}}
\def \beqa {\begin{eqnarray}}
\def \eeqa {\end{eqnarray}}

\newcommand{\mbold}[1]{\mbox{\boldmath$#1$}}
\documentstyle[preprint,aps]{revtex}

\begin{document}
% >>>>>>>>>>>>>>>>>>>>>>>>>>>>>>>>>>>>>>>>>>>>>>>>>>>>>>>>>>>>>>>>>>>
% TITLE AND AUTHORS.
%
\begin{center}
{\large\bf  ON THE THOMAS-FERMI APPROXIMATION FOR BOSE
CONDENSATES IN TRAPS}
\\[2.cm]
\centerline {P. Schuck$^{1}$
and X. Vi\~nas$^{2}$}

{\it $^1$Institut des Sciences Nucl\'eaires,} \\
{\it Universit\'e Joseph Fourier, CNRS--IN{\sl 2}P{\sl 3},} \\
{\it {\sl 53} Avenue des Martyrs, F-{\sl 38026} Grenoble-C\'edex,
     France}

{\it $^2$Departament d'Estructura i Constituents de la Mat\`eria, \\
Facultat de F\'{\i}sica, Universitat de Barcelona,} \\
{\it Diagonal {\sl 647}, E-{\sl 08028} Barcelona, Spain} \\[1.0cm]
\vspace{0.5cm}
PACS number(s) 03.75.Fi, 05.30.Jp
\end{center}
% >>>>>>>>>>>>>>>>>>>>>>>>>>>>>>>>>>>>>>>>>>>>>>>>>>>>>>>>>>>>>>>>>>>
% ABSTRACT.
%
\vspace{1.5cm}
\begin{abstract}
Thomas-Fermi theory for Bose condesates in inhomogeneous traps is
revisited. The phase-space distribution function in the Thomas-Fermi
limit is $f_0(\mbold{R},\mbold{p})$  $\alpha$  $\delta(\mu - H_{cl})$
where $H_{cl}$ is the classical counterpart of the self-consistent
Gross-Pitaevskii Hamiltonian. No assumption on the large N-limit is
introduced and, e.g the kinetic energy is found to be in good
agreement with the quantal results even for low and intermediate
particle numbers N. The attractive case yields conclusive results as
well.
\end{abstract}
\pagebreak

\section{Introduction}

The recent discovery of the Bose-Einstein condensation of magnetically
trapped atoms has spurred a huge amount of theoretical investigations.
Most of them are based on the Gross-Pitaevskii equation (GPE)
\cite{Pitae} which is the mean-field equation for the condensate
wave function (order parameter).
The experimental conditions are such that the atomic
gas is at very low density and therefore the mean-field approximation
gives indeed excellent results \cite{Dalf1,Dalf2,Dalf3,Park}. Since the
number N of atoms
involved is generally large, it is natural that also the Thomas-Fermi
(TF) approximation is applied quite extensively. This has the
advantage of yielding in most cases explicit analytical results of
great physical transparency. However, with respect to Fermi
statitstics, the TF-approach to Bose-Einstein condensation shows
some peculiarities which in the past have, in our opinion, not
fully been born out. It is the purpose of the present paper to further
elaborate on the TF-approach to inhomogeneous Bose systems. We
deliberately restrict ourselves here to the TF-limit of the GPE at
zero temperature. Finite temperature as well as more elaborated
theories like the Bogoliubov approach may be the subject of future
work.

In detail the paper is orgasnized as follows. In Section 2 the theoretical
aspects of the TF approximation to the GPE are presented in detail. In
Section 3 the numerical results obtained with the TF method are compared with
the ones coming from the GPE. The last Section is
devoted to discussions and an outlook.

\section{Gross-Pitaevskii equation. Thomas-Fermi limit}

As mentioned in the introduction, the basic equation for Bose
condensed atoms confined by magnetic traps is, in the low density
limit, given by the GPE for the wave function of the condensate:
\beq
( - \frac{\hbar^2}{2m} \Delta + V_{ex} + g {\vert \psi \vert}^2 )
\psi = \mu \psi
\label{eq1}\eeq
where $V_{ex}$ is the external potential which for simplicity we have
considered to be a spherical harmonic oscillator (for non-spherical
geometry see remarks at the end of the paper). The coupling constant
is given by $g= 4 \pi \hbar^2 a/m$ with $m$ the atomic mass and $a$
the scattering length. The chemical potential $\mu$ is identical with
the lowest eigenvalue of the self-consistent potential
\beq
 V = V_{ex} +
g {\vert \psi \vert}^2.
\label{eq1b}\nonumber\eeq
It is useful to note that eq.(\ref{eq1}) can also be rewritten as an
equation for the density $\rho = {\vert \psi \vert}^2$ :
\beq
\frac{\hbar^2}{2 m} \frac{1}{4} \bigg[ \frac{(\nabla \rho)^2}{\rho^2}
- 2 \frac{\Delta \rho}{\rho} \bigg] + V_{ex} + g \rho -\mu = 0
\label{eq1a}\eeq
In the large $N$ limit one can drop in (\ref{eq1}) the kinetic
energy or, equivalently, in (\ref{eq1a}) the gradients terms of the
density. This leads to $\rho =(\mu - V_{ex})/g$ what is known in
literature \cite{Dalf2,Dalf3,Park} as the TF solution of the GPE.
However, for moderate particle numbers the kinetic
energy is not negligeable and there is no reason for dropping it in
the TF limit. Also in the attractive case the kinetic energy is of
crucial importance to avoid collapse.

As it is well known from the case of the Fermi systems \cite{Ring},
the TF-approximation is based on the assumption of a slowly varying
potential so that its gradients can be neglected to lowest order.
The TF approximation of the density matrix corresponding to a single
wave function is, as a matter of fact, well known \cite{R5} and
given by:
\beq
\lim_{\hbar \to 0} {\left[ \psi(\mbold{r}) \psi(\mbold{r^\prime})
\right]}_W
=  N c \delta (\mu - H_c) + {\cal{O}}(\hbar^2)
\label{eq2}\eeq
where the index $"W"$ stands for the Wigner transform
\cite{Ring} and $H_c$ is the classical hamiltonian:
$H_c = \frac{p^2}{2m} + V_{ex} (\mbold{R}) + g \rho(\mbold{R})$.
The constant $c(\mu)$ of dimension of an inverse
of energy is determined from the condition that the wave function must
be normalized leading to :
\beq
\frac{1}{c} = \int \frac{d \mbold{R} d\mbold{p}}{(2 \pi \hbar)^3}
\delta (\mu - H_c)
\label{eq3}\eeq
As a consequence the chemical potential $\mu$ must be determined from
an independent quantization condition (see below). We will, however,
see that in the limit $N \to \infty$ the expression for $\mu$
coincides with the usual.

It is clear from (\ref{eq1a}) that the kinetic energy has been properly
included (as usual in TF theory).
One way to understand expression
(\ref{eq2}) is to write eq.({\ref{eq1}) in the form $(\mu -
H)\hat{\rho}=0$ with $\hat{\rho}=\vert \psi \rangle \langle \psi
\vert$
the density matrix. Wigner transforming this equation and remembering
that to lowest order in $\hbar$ the Wigner transform of a product of
operators is the product of their Wigner transforms \cite{Ring}, one
arrives with $x \delta(x) = 0$ at eq.(\ref{eq2}).

\subsection{Self-consistent solution. Repulsive case ($g > 0$)}

Let us first consider the solution of the self-consitent
problem at the TF level defined by eq.(\ref{eq2}) for
the repulsive case i.e $g > 0$. From (\ref{eq2}) we obtain
for the density:
\beq
\rho(\mbold{R}) = \int \frac{d\mbold{p}}{( 2 \pi \hbar)^3} f_0(\mbold{R},
\mbold{p}) =
\frac{N c m}{2 \pi^2 \hbar^3} p_0 (\mbold{R})
\label{eq4}\eeq
where the local momentum is given by :
\beq
p_0 (\mbold{R}) = \sqrt{2m (\mu - V_{ex} - g \rho)}
\label{eq5}\eeq
The self-consistency between eqs.(\ref{eq4},\ref{eq5}) is easy to
solve analytically and we obtain :
\beq
\rho = - \frac{K g N^2}{2} + \sqrt { (\frac{K g N^2}{2})^2 + K N^2
(\mu - V_{ex})}
\label{eq6}\eeq
where
\beq
K = \frac{2m}{\hbar^2} (\frac{c m}{2 \pi^2 \hbar^2})^2
\label{eq6a} \eeq
It is to be noted that $\rho(\mbold{R})$ is defined only within the
classical region limited by $\mu- V_{ex}(\mbold{R})=0$. It is
straightforward to expand $\rho$ in the repulsive case $g>0$ for large
values of $N$ :
\beq
\rho \approx \frac{1}{g}(\mu - V_{ex}) - \frac{1}{K g^3 N^2} (\mu -
V_{ex})^2 + .....
\label{eq7}\eeq
It is satisfying to see that to leading order one recovers the
result corresponding to the total neglect of  kinetic energy in
(\ref{eq1}) (see introduction).

The normalization is directly determined from eq.(\ref{eq3})
\beq
\frac{1}{c}= \frac{m}{2 \pi^2 \hbar^3} \int d \mbold{R} p_0\mbold{R})
\label{eq8}\eeq
or equivalently
\beq
N = \int d \mbold{R} \rho(\mbold{R})
\label{eq9}\eeq
Explicitly one obtains from this equation:
\beqa
1 & = & 4 \pi \bigg( \frac{2}{m \omega^2}\bigg)^{3/2} \bigg\{ - \frac
{5 K g N}{48} \mu^{3/2} - \frac{K^2 g^3 N^3}{64} \mu^{1/2}
\nonumber\\
&+& \mbox{} \frac{\sqrt{K}}{8} \bigg(\frac{K g^2 N^2}{4} + \mu
\bigg)^2
\arcsin{\sqrt{\frac{\mu}{\frac{K g^2 N^2}{4} + \mu}}}\bigg\}
\label{eq10}\eeqa
Using in (\ref{eq9}) the asymptotic expansion (\ref{eq7}) yields
\beq
1 = 4 \pi \big( \frac{2}{m \omega^2}\bigg)^{3/2}  \frac{2}{15}
\frac{\mu^{5/2}}{g N} \bigg( 1 - \frac{4}{7} \frac{\mu}{K g^2 N^2}
\bigg)
\label{eq11}\eeq
which also can directly be derived in expanding (\ref{eq10}).

From Eqs.(\ref{eq10}) or (\ref{eq11}) we can determine the normalization
constant $c$ as a function of $\mu$. From (\ref{eq11}) we see that to
lowest order the normalization constant $c$ drops out
and thus in this limit the chemical potential is, as
usual, determined by the particle number condition.
However, as we will see, via the quantization condition
$\mu$ depends on $c$ and thus we can consider (\ref{eq10}) as
determining the normalization in any case.

\subsection{Chemical potential and quantization}

The semiclassical density matrix (\ref{eq2}) corresponds to a
single wave function. In such a case the
energy must be determined independently from a quantization
condition. This is in fact well known \cite{KCM}. Formally
in our case this is necessary, since we have two open constants $\mu$ and
K. The equation for the chemical potential $\mu = dE/dN$ is also of
no help, since it is equivalent to the particle number condition
(\ref{eq9})
The standard semiclassical quantization procedure is given by the
WKB method. However, in order to have a more explicit formula, we here also
apply a slightly simpler method, aplicable to the lowest state in a
single particle potential \cite{Carbo}. To this end we calculate the
smooth accumulated level density (number of states) in TF
approximation:
\beq
N^{TF} (E) = \int \frac{d \mbold{R} d\mbold{p}}{(2 \pi \hbar)^3}
\Theta (E - H_c)
\label{eq12}\eeq
For a spherical harmonic oscillator (H.O.) this gives:
\beq
N^{TF}_{H.O.} = \frac{1}{6} {\bigg(\frac{E}{\hbar \omega}\bigg)}^3
\label{eq13}\eeq
Taking for $E$ in (\ref{eq13}) the H.O. eigenvalues
\beq
E  \to E_K = (K + \frac{3}{2}) \hbar \omega
\label{eq14}\eeq
with $K = 2 n + l$ and inserting (\ref{eq13},\ref{eq14}) in the
left hand side of (\ref{eq12}) yields a semiclassical quantization
rule
which becomes exact in the 3-D spherical H.O. case. It represents an
approximate quantization relation for an arbitrary potential where the
quantized energies very well reproduce the centroid of major shells.
This has been tested numerically on a potential of Woods-Saxon type
of nuclear dimensions \cite{Carbo}. It is evident that in the 1-D
case the same procedure leads to the exact WKB quantization
rule. For the 3-D case this modified quantization prescription is
slightly less accurate than WKB for the lowest eigenvalue
but has the advantage to be easier and to be readly applicable also to
the deformed case \cite{Carbo}. In the present problem the eigenvalue
$\mu$ is then determined by:
\beqa
\frac{27}{8} & = & \frac{1}{\pi^2 \hbar^3} \int d\mbold{R} {p_0}^3
(\mbold{R}) \nonumber\\ & = & \mbox{}
\frac{32}{\pi \hbar^3} \bigg\{
- \frac{27}{80}\bigg(\frac{K g^2 N^2}{4}\bigg)^{1/2} \mu^{5/2}
- \frac{19}{24}\bigg(\frac{K g^2 N^2}{4}\bigg)^{3/2} \mu^{3/2}
- \frac{7}{16}\bigg(\frac{K g^2 N^2}{4}\bigg)^{5/2} \mu^{1/2}
\nonumber\\ &+& \mbox{}
\frac{3}{8} \frac{K g^2 N^2}{4}
\bigg(\frac{K g^2 N^2}{4} + \mu\bigg)^2
\arcsin{\sqrt{\frac{\mu}{\frac{K g^2 N^2}{4} + \mu}}}
\nonumber\\ &+& \mbox{}
\frac{1}{16} \bigg(\frac{K g^2 N^2}{4} + \mu\bigg)^3
\arcsin{\sqrt{\frac{\mu}{\frac{K g^2 N^2}{4} + \mu}}}\bigg\}
\label{eq15}\eeqa
where we have used $p_0(\mbold{R})$ from (\ref{eq5}) with (\ref{eq6}).
To second order we obtain from (\ref{eq15}):
\beq
\frac{27}{8} =  \frac{4}{\pi \hbar^3}
\bigg( \sqrt{\frac{2 m}{K}} \frac{1}{g N}\bigg)^3
\bigg(\frac{2}{m \omega^2}\bigg)^{3/2}  \mu^{9/2}
\frac{16}{315} \bigg( 1 - \frac{48}{11} \frac{\mu}{K g^2 N^2}\bigg)
\label{eq16}\eeq
To leading order in the large $N$ limit we obtain from
(\ref{eq10},\ref{eq11}) and (\ref{eq15},\ref{eq16}):
\beqa
\mu_0 & = & \bigg(\frac{15}{8 \pi}\bigg)^{2/5} \bigg(\frac{m
\omega^2}{2}\bigg)^{3/5} \bigg(g N\bigg)^{2/5}
\nonumber\\
K_0 & = & \bigg( \frac{4096}{8505 \pi \hbar^3 \omega^3} \bigg)^{2/3}
\frac{1}{(g N)^2} {\mu_0}^3
\label{eq17}\eeqa
This completes the solution in the large $N$ limit.

For later comparison let us also give the standard WKB quantization
rule \cite{Mesh} which we want to evaluate to leading order :
\beq
\frac{\pi}{2} = \sqrt{\frac{2 m}{\hbar^2}}\int^{r_2}_{r_1} dr
\bigg[ \mu - V(r) -\frac {\hbar^2}{2 m} \frac{1/4}{r^2}\bigg]
\label{eq16a}\eeq
with (see eq.(\ref{eq7}))
\beq
\mu - V(r) = \frac{1}{K g^2 N^2} \big(\mu - V_{ex}(r) \big)^2
\label{eq16b}\eeq
The classical turning points $r_{1}$ and $r_{2}$ are determined from the
solution of the cubic equation:
\beq
\frac{1}{\sqrt{K g^2 N^2}} (\mu - V_{ex}) = - \frac{1}{2}
\sqrt{\frac{\hbar^2}{2 m}} \frac{1}{r}
\label{eq16c}\eeq

With (\ref{eq16b}) and (\ref{eq16c}) eq.(\ref{eq16a}) can
be solved for $\mu$.

\subsection{Kinetic energy}
One of the main difficulties with the standard $N \to \infty$ TF limit
treated in the literature (see introduction) consists in the inability to
calculate the kinetic energy \cite{Dalf2}. In our approach this does
not cause any particular problem and one directly obtains
\beq
E_{kin} = \int \frac{d\mbold{R} d\mbold{p}}{(2 \pi \hbar)^3}
\frac{p^2}{2 m} f_0\big(\mbold{R},\mbold{p}\big) = \frac{N c}{4
\pi^2 \hbar^3} \int d\mbold{R} {p_0}^3 (\mbold{R})
\label{eq18}\eeq
For example to lowest order we obtain from (\ref{eq5},\ref{eq7})
\beq
p_0 = \frac{1}{\sqrt{K_0} g N} \big(\mu_0 - V_{ex}) \Theta \big(\mu_0
- V_{ex})
\label{eq19}\eeq
what yields for the kinetic energy per particle, using (\ref{eq17})
\beq
\frac{{E^{(0)}_{kin}}_0}{N} = \frac{27}{32} \frac{2 \pi^2 \hbar^3}{m}
\sqrt{\frac{K_0}{2 m}} = \frac{27}{32} c_0
\label{eq20}\eeq
This simple result must be contrasted with the usual statment that in
the $N \to \infty$ TF limit the kinetic energy can not be evaluated,
since it diverges \cite{Dalf2}.

Indeed one can write the kinetic energy as:
\beq
E_{kin} = \frac{\hbar^2}{2 m} \int d\mbold{R} {\vert \nabla \psi
\vert}^2
\label{eq21}\eeq
In the $N \to \infty$ limit we have:
\beq
 \psi_{N \to \infty} = \sqrt{\frac{1}{g} \big(\mu_0 - V_{ex}\big)}
\label{eq22}\eeq
and one can readily verify that with (\ref{eq22}) $E_{kin}$ of
of (\ref{eq21}) diverges
logaritmically. This result obviously is in contradiction with
(\ref{eq20}) and we shortly want to elucidate the underlying reason. To
this end we first rewrite (\ref{eq21}) in a different but obviously
equivalent way:
\beq
E_{kin} = \frac{\hbar^2}{2 m} \int d\mbold{R} {\vert \nabla \psi
\vert}^2 = \int \frac{d\mbold{R} d\mbold{p}}{(2 \pi \hbar)^3}
\frac{p^2}{2 m} \tilde{f_0} \big(\mbold{R},\mbold{p}\big)
\label{eq23}\eeq
with $\tilde{f_0}$ given by the Wigner transform of the density matrix
corresponding to (\ref{eq22}):
\beq
\tilde{f_0}\big(\mbold{R},\mbold{p}\big) = \int d\mbold{s} e^{-i
\mbold{p}\mbold{s}/\hbar} \psi_{N \to
\infty}\big(\mbold{R}+\frac{\mbold{s}}{2}\big) \psi_{N \to \infty}
\big(\mbold{R}-\frac{\mbold{s}}{2}\big)
\label{eq24}\eeq
Since $\tilde{f_0} \ne f_0$ we argue that (\ref{eq24}) is not the
correct $\hbar \to 0$ limit of the distribution function because it is
not solution of the $\hbar \to 0$ limit of the Schr\"odinger equation
(\ref{eq1})
\beq
(H_c - \mu) f_0 = 0
\label{eq25}\eeq
Only (\ref{eq2}) is the correct solution of this equation which yields
for large $N$:
\beq
f_{0_{N \to \infty}} = N c \delta \bigg(\frac{1}{K_0 g^2
N^2}{\big(\mu_0 - V_{ex}\big)}^2 - \frac{p^2}{2 m}\bigg)
\label{eq26}\eeq
One checks that upon projection onto r-space (\ref{eq26}) gives the
correct lowest order expression for the density (see eq.(\ref{eq7})).
Therefore both Wigner functions (\ref{eq24}) and (\ref{eq26}) yield
the same leading order density. However, in spite of being a very
suggestive non-local generalization of the lowest order local density
expression, eq.(\ref{eq24}) has to be rejected on the above given
grounds and the divergency of (\ref{eq21}) is an artifact. On the
contrary the lowest order contribution to the kinetic energy is given
by (\ref{eq20}). Via (\ref{eq11},\ref{eq16}) it is straightforward to
calculate the next to leading order correction to (\ref{eq20}). It
should, however, be remembered that $1/N$ correction do not go in
parallel with powers in $\hbar$ and that $1/N$ corrections
also can come from $\hbar^2$ corrections to (\ref{eq2}) which involve
second order gradients of the potential. In any case the
Wigner-Kirkwood expansion of the density matrix is an asymptotic
expansion which in no way can recover the nonanalytic behavior in
$\hbar$ of the quantal solution. In the present problem the
nonanalyticity in $\hbar$ of the quantal solution entails a nonanalytic
behavior in
$1/N$ (see eq.(16) of ref.\cite{Dalf2}) and therefore a WK-expansion
can never recover the quantal behavior in $1/N$. It is well known that
an asymptotic expansion has to be stopped at a point where the
difference to the exact solution is minimal. Afterwards the expansion
starts to diverge again.
In this work we do not intend to develope a systematic expansion
simultaneously in $\hbar$ and $1/N$. We rather want to give a complete
solution to lowest order in $\hbar$, i.e. on the TF level.

\subsection{The attractive case ($g < 0$)}
It seems that recently, Bose-Einstein condensation has been observed
also for the case of negative scattering length ($^{11}Li$ atoms )
\cite{Brad}.

For $g<0$ the Gross-Pitaevskii approach leads to metastability for particle
numbers $N \le 1400$ \cite{Dalf1}. For large particle numbers the
system
collapses. For the attractive case ($a < 0$) the correct treatment of
the kinetic energy is crucial in the TF-limit, since otherwise no
stability can be achieved. Formally the TF solution for the density
is the same as in (\ref{eq6}) with, however, the sign of the first
member reversed :
\beq
\rho = \frac{K \vert g \vert N^2}{2} + \sqrt { (\frac{K g N^2}{2})^2 +
K N^2
(\mu - V_{ex})}
\label{eq27}\eeq
Contrary to the repulsive case no large $N$ expansion is
possible here. Therefore the TF solution has to be considered
in full. In the next section comparison with quantal results will be
given.

\section{Numerical Results}
In this section we proceed to a detailed numerical comparison of the
semiclassical approximations with the exact quantum mechanical
results. Along this section energies and lengths are given in
harmonic oscillator units : $\hbar \omega$ and
$a_{HO}=\sqrt{\hbar/2 m \omega}$, respectively. First in Table
1 we present the chemical potential
($\mu$) and
the kinetic ($e_{kin}$), harmonic oscillator
($e_{HO}$) and the self-interaction ($e_{\rho}$) energies per particle
calculated quantally and in the full Thomas-Fermi approximation
(\ref{eq6},\ref{eq10},\ref{eq15}) as a function of the number of
atoms enclosed in the trap. We have considered $Cs$ atoms (as was
done in Ref. \cite{Edw}), the frequency of the harmonic oscillator
has been chosen to be $\omega = 20 \pi s^{-1}$ and the scattering
length to be $a = 3.2$x$10^{-9} m$.

In Table 2 we present the results for the chemical potential and the
kinetic energy per particle number beyond 20000. In addition
to the quantum mechanical and the full Thomas-Fermi results,
we also include the results for the large $N$ limit
(\ref{eq20}) and those obtained using the WKB
quantization rule (in the large $N$ limit). Notice that in the large N
limit, the WKB chemical potential coincides with the TF one and the
kinetic energy is given also by (\ref{eq20}) but with $c_0$
replaced by the one calculated from $K$ via eq.(\ref{eq16a}).
From Tables 1, 2 we
see that for instance the results of the full TF solution are in very
satisfaying agreement with the quantal results over the whole range of
particle numbers.

The numbers presented in Table 2 indicate that the asymptotic values
of the chemical potential and the kinetic energy are obtained only for
very large number of particles ($N \simeq 10^5-10^6$). We also realise
that the WKB quantization rule yields quite similar results though in
fact slightly worse ones than our simpler TF-quantization rule
(\ref{eq15},\ref{eq16}). Though globally the semiclassical results of
Tables 1 and 2 are quite satisfactory, one nevertheless remarks some
unexpected features. For instance the kinetic energy in the TF
approximation is larger than the exact values for small numbers of
particles whereas it undershoots the quantum values quite considerably
in the large $N$ limit. We will come back to a more detailed analysis
of this behavior in the discussion section.

Next let us compare in Fig.1 and 2 the densities (normalized to
unity) in TF-approximation
and calculated exactly for small (200) and large (200000) particle
numbers. As expected, the TF densities almost agree with the quantal
ones for very large particle numbers. In view of the still quite
reasonable expectation values shown in Table 1 for N=200, the strong
deviation of the TF density from the quantal result is somewhat a
surprise. However, one always should remember that the TF-solution for
the densities is to be understood as a distribution (see
eq.\ref{eq2}) which
for expectation values of "slowly varying" operators can still yield
very reasonable values in spite of the fact that the detailed shape
may only be
a charicature of the exact one.

In Figs.3 and 4 we show the self-consistent potentials $V = V_{ex} + g
\rho$ corresponding to the densities of Figs.1 and 2. Not astonishingly
$V$ deviates from the harmonic oscillator only slightly for $N=200$,
both quantally and semiclassically. On the contrary for $N=200000$,
the potential $V$ deviates strongly from $V_{ex}$ being practically
a constant equal to $\mu$ up to the classical turning point from where
the harmonic oscillator takes over quite  abruptly. Again, both quantal and TF-solution are in close agreement. Fig.4
also teaches us why the TF-approximation (\ref{eq2}) to the quantal
distribution function is very good for large N. The distribution
function corresponds to a wavefunction with very large energy $\mu$.
In phase space it therefore is very much concentrated around the
surface of the
hypersphere with radius $\mu$.

Let us now present the attractive case for the same atoms and
external potential with ,however, the scattering length $a=-1.0$x$10^{-9}m$. In Table 3 we
again show chemical potential and kinetic, harmonic oscillator
and self-interaction energies per particle  as a function of the
particle number in
TF and quantal calculation. For small particle numbers ($N \le 1000$)
the agreement of TF with the quantal case is of similar quality as in
the repulsive case. However for $N \ge 1000$ the agreement quickly
deteriorates, indicating that the whole mean field approximation
breaks down. Indeed even quantally the solution of the GPE (\ref{eq1})
becomes unstable for $N>1500$ for $Cs$ atoms.

In Figs.5, 6, 7 and 8 we also show the densities (normalized to
unity) and self-consistent
potentials for the particle numbers $N=250$ and $N=1500$. We see that,
whereas the case $N=250$ is not dissimilar to the corresponding one
with $a>0$, for $N=1500$ the situation becomes quite unfavorable for the
TF approximation. This is for instance manifest in looking at the
graph for the densities. In the attractive case TF and quantal
solutions diverge with increasing $N$ whereas in the repulsive case
they converge.

In Figs.9 and 10 we plot the kinetic energy density per
particle ($\tau/N$) calculated
quantally and in the TF-approximation for $N=200000$ in the
repulsive case and for $N=250$ in the attractive case. In these Figures
the quantal kinetic energy density is given by:
\beq
\tau = {\vert \nabla \psi \vert}^2 - \frac{1}{4} \Delta \rho =
= \frac{1}{4} \bigg[ \frac{(\nabla \rho)^2}{\rho} - \Delta \rho \bigg]
\label{eq27a}\eeq
in order to compare with the TF one according to Ref.\cite{Ring}

For large number of particles, when the density profile has a
relatively flat region at the interior (see Fig.2),
the quantal kinetic energy density is peaked at the
surface (see eq.(\ref{eq27a})) whereas the TF one is rather a bulk term
(see eq.(\ref{eq18})). Inspite of the rather different form of the
quantal and TF kinetic energy densities in this case, the corresponding integrals
are in good agreement (see Tables II). This fact points
again to the distribution character of the TF-kinetic energy density.
However, if the number of particles is small, the quantal and TF kinetic
energy density profiles are quite similar. This is due to the fact that
the particle density fall-off abruptly from $R=0$ (see Fig.5) and
consequently its derivatives contribute in all the range of $R$.

\section{Discussion and Outlook}
In the preceding sections we have derived the Thomas-Fermi
approximation i.e. the $\hbar \to 0$ limit of the density matrix
corresponding to the wavefunction of the Bose condensate of atoms
confined by magnetic traps. We have pointed out some misconceptions on
this point which appeared in the past in the literature which for
instance prevented the direct calculation of the kinetic energy in the
large $N$ limit. On the contrary with our Thomas-Fermi approach the
evaluation of the kinetic energy causes no problem and the results are globally
in quite satisfactory agreement with the quantal solution of the
Gross-Pitaevskii equation. However from Table 2 we see that the
kinetic energy in the TF limit does not have the correct asymptotic
behavior as a function of particle number. As was shown by Pitaevskii
and Stringari \cite{Dalf2}, this is due to nonanalytic (logarithmic)
corrections which can not be accounted for by a pure TF approach and
needs a partial resummation of all orders in $\hbar$. However, for particle
numbers where the kinetic energy represents a significant fraction of
the total energy the TF expression yields very satisfying results for
$E_{kin}$. This example shows again that the
semiclassical approximations are a powerful tool but not devoid of
subtleties and pitfalls. As a matter of fact also in this paper we,
for simplicity, avoided to develope the full complexity of the theory.
One major simplification resides in the fact that we assume a
spherical trap. This results in an isotropic momentum distribution
$f_0$ $\alpha$ $\delta (\mu - H_c)$ where $H_c = \frac{p^2}{2m}+V$ is the
classical Hamiltonian. Deforming the trap leads to a non trivial
modification of the theory, since squeezing the condensate wavefuntion
in one direction and relaxing in the other entails in turn a
deformation of the momentum distribution which is opposite to the
spatial one, i.e. momenta are strongest in the squeezed direction and
lowest in the long direction of the deformation \cite{Jen}.
Our TF approach can also be useful for the evaluation of collective
excitations of droplets of small or intermediate sizes.
Such a situation is
in fact well known from the zero sound giant quadrupole vibrations of
finite nuclei where to first approximation the restoring force of the
vibration is given by the energy stored in the deformation of the
Fermi sphere \cite{Ring}. On the other hand the deformation of the
momentum distribution of condensed atoms has also been revealed
experimentally in directly measuring the momentum distribution of the
expanding particles, once the deformed trap has been turned off. The
detailed determination of the anisotropy of the momenta (which may be
position dependent) is theoretically a not completely trivial task in
the general case and we will elaborate on this in future work. In the
present case, however, there exists an evident first guess of the
momentum deformation which results from a scaling argument of the
harmonic oscillator coordinates. Assuming a prolate quadrupole
deformation in the $z$-direction we have to replace the classical
hamiltonian in (\ref{eq2}) by :
\beq
\tilde H_{c} = \frac{1}{2m}\bigg[
\frac{{\omega_0}^2}{{\omega_{\perp}}^2}({p_x}^2+{p_y}^2) +
\frac{{\omega_0}^2}{{\omega_z}^2} {p_z}^2 \bigg]
+ V \bigg( \frac{\omega_{\perp}}{\omega_0} x,
\frac{\omega_{\perp}}{\omega_0} y, \frac{\omega_z}{\omega_0}z \bigg)
\label{eq28}\eeq
where the ratios $\frac{\omega_z}{\omega_0}$ and
$\frac{\omega_{\perp}}{\omega_0}$ are the frequency relations in $z$
and $x,y$ with respect to the spherical case ($\omega_0$). From
(\ref{eq28}) one easily calculates the so-called aspect ratio in the
TF-approximation
\beq
\sqrt{\frac{{p_z}^2}{{p_x}^2}}  = \frac{\omega_{\perp}}{\omega_z}
\label{eq29}\eeq
a result which has been given previously \cite{Dalf2}. One other
important consequence of the momentum deformation is that with
(\ref{eq28}) the moment of inertia of the condensate becomes equal
to the irrotational flow value \cite{String}.
On the contrary using (\ref{eq2}) with the
isotropic momentum distribution the rigid momentum of inertia
results. Consequently the deformed case needs more detailed studies
which we reserve to future work. It is also evident that the
present TF approach can be extended to finite temperature and
to the Bogoliubov theory.

Another interesting subject of a more formal aspect is the evaluation
of the $\hbar$-correction to the present lowest order theory. In
principle this can easily be performed in posing in (\ref{eq1a})
$\rho = \rho_0 + {\hbar}^2 \rho_2$ and $\mu = \mu_0 + {\hbar}^2 \mu_2$
and properly sorting out different powers in $\hbar$. However the
proper elimination of divergencies and handling the normalization
({\ref{eq9}) and quantization (\ref{eq15}) are slightly subtle
problems.

Investigations on the above mentioned directions are in
progress.

\section*{Acknowledgments}
\noindent

We want to thank S.Stringari for very useful
discussions and L.P.Pitaevskii for his interest in this work. One of us (X.V) also acknowledges financial support
from DGCYT (Spain) under grant PB95-1249 and from the DGR (Catalonia)
under grant GR94-1022.
\pagebreak
% >>>>>>>>>>>>>>>>>>>>>>>>>>>>>>>>>>>>>>>>>>>>>>>>>>>>>>>>>>>>>>>>>>>
% REFERENCES.
%

\pagebreak
% >>>>>>>>>>>>>>>>>>>>>>>>>>>>>>>>>>>>>>>>>>>>>>>>>>>>>>>>>>>>>>>>>>
% THE TABLES.
%
\section*{Tables}
\noindent

TABLE I. Chemical potential ($\mu$), kinetic energy ($e_{kin}$),
harmonic oscillator energy ($e_{HO}$) and self-interaction energy
($e_{\rho}$) per particle in harmonic oscillator units
($\hbar \omega$) calculated quantally
(QM) and
in the Thomas-Fermi approaximation (TF) for several numbers of atoms
in the traps. The frequency of the harmonic oscillator is $\omega$= 20
$\pi$ s$^{-1}$ and the scattering length is a= 3.2x10$^{-9}$ m.
\vspace{1cm}

TABLE II. Chemical potential ($\mu$) and kinetic energy
($e_{kin}$) per particle for large number of atoms in the trap (N).
The chemical
potential is calculated quantally (QM), with the full Thomas-Fermi
approach (TF) and with the asymptotic formula for large N (TF$_{N \to
\infty}$). The kinetic energy is obtained quantally, with the exact
TF approximation, with the asymptotic TF for large N and using the WKB
quantization in the limit of large number of atoms. The frequency of
the harmonic oscillator and the scattering lenght are the same as in
Table I.

TABLE III. The same as in Table I but with a scattering length a=
-1.0x10$^{-9}$ m.
\vspace{1cm}

\pagebreak
 \begin{center}
   TABLE I
 \end{center}

\begin{center}  \small
\begin{tabular}{l c c c c c c c c}
\hline
$N$ & $\mu$(QM) & $\mu$(TF) & $e_{kin}$(QM) & $e_{kin}$(TF)
  & $e_{HO}$(QM) & $e_{HO}$(TF)
  & $e_{\rho}$(QM) & $e_{\rho}$(TF)  \\
\hline
200
  & 1.688 & 1.642 & 0.696 & 0.700 & 0.811 & 0.804 & 0.080 & 0.069 \\
400
  & 1.806 & 1.766 & 0.654 & 0.661 & 0.865 & 0.851 & 0.144 & 0.127 \\
600
  & 1.927 & 1.877 & 0.622 & 0.630 & 0.912 & 0.894 & 0.196 & 0.177 \\
800
  & 2.036 & 1.978 & 0.597 & 0.603 & 0.955 & 0.934 & 0.242 & 0.220 \\
1000
  & 2.134 & 2.071 & 0.575 & 0.581 & 0.944 & 0.970 & 0.282 & 0.260 \\
1200
  & 2.225 & 2.157 & 0.557 & 0.561 & 1.031 & 1.005 & 0.319 & 0.296 \\
1400
  & 2.310 & 2.238 & 0.541 & 0.544 & 1.065 & 1.037 & 0.352 & 0.329 \\
1600
  & 2.389 & 2.315 & 0.528 & 0.528 & 1.065 & 1.068 & 0.382 & 0.359 \\
1800
  & 2.464 & 2.388 & 0.515 & 0.514 & 1.127 & 1.097 & 0.411 & 0.388 \\
2000
  & 2.535 & 2.457 & 0.503 & 0.502 & 1.158 & 1.125 & 0.437 & 0.415 \\
4000
  & 3.112 & 3.025 & 0.431 & 0.417 & 1.395 & 1.356 & 0.643 & 0.626 \\
6000
  & 3.550 & 3.461 & 0.390 & 0.369 & 1.577 & 1.536 & 0.792 & 0.778 \\
8000
  & 3.914 & 3.825 & 0.363 & 0.336 & 1.729 & 1.687 & 0.911 & 0.901 \\
10000
  & 4.231 & 4.142 & 0.343 & 0.312 & 1.862 & 1.820 & 1.013 & 1.005 \\
12000
  & 4.513 & 4.426 & 0.327 & 0.293 & 1.981 & 1.939 & 1.103 & 1.097 \\
14000
  & 4.770 & 4.684 & 0.314 & 0.277 & 2.089 & 2.047 & 1.184 & 1.180 \\
16000
  & 5.007 & 4.921 & 0.303 & 0.265 & 2.189 & 2.147 & 1.258 & 1.255 \\
18000
  & 5.228 & 5.143 & 0.294 & 0.254 & 2.282 & 2.240 & 1.326 & 1.324 \\
20000
  & 5.435 & 5.350 & 0.285 & 0.244 & 2.369 & 2.328 & 1.390 & 1.389 \\
\hline
\end{tabular}
\end{center}
\pagebreak

 \begin{center}
   TABLE II
 \end{center}

\begin{center}  \small
\begin{tabular}{l c c c c c c c}
\hline
$N$ & $\mu$(QM) & $\mu$(TF) & $\mu$(TF$_{N \to \infty}$)
  & $e_{kin}$(QM) & $e_{kin}$(TF) & $e_{kin}$(TF$_{N \to \infty}$) &
$e_{kin}$(WKB$_{N \to \infty}$) \\
\hline
20000
  & 5.435 & 5.350 & 5.196 & 0.285 & 0.244 & 0.256 & 0.238 \\
30000
  & 6.322 & 6.242 & 6.111 & 0.255 & 0.210 & 0.218 & 0.202 \\
40000
  & 7.051 & 6.973 & 6.856 & 0.236 & 0.187 & 0.194 & 0.180 \\
50000
  & 7.677 & 7.603 & 7.496 & 0.222 & 0.173 & 0.177 & 0.165 \\
\\
100000
  &10.231 & 9.972 & 9.891 & 0.182 & 0.133 & 0.134 & 0.125 \\
150000
  &11.763 &11.701 &11.633 & 0.162 & 0.113 & 0.114 & 0.106 \\
200000
  &13.170 &13.112 &13.051 & 0.149 & 0.101 & 0.102 & 0.095 \\
250000
  &14.381 &14.326 &14.270 & 0.140 & 0.093 & 0.093 & 0.087 \\
\hline
\end{tabular}
\end{center}

 \begin{center}
   TABLE III
 \end{center}

\begin{center}  \small
\begin{tabular}{l c c c c c c c c}
\hline
 & QM & TF & QM & TF & QM & TF & QM & TF \\
\hline
$N$
  & 250 & 250 & 500 & 500 & 1000 & 1000 & 1500 & 1500  \\
$\mu$
  & 1.424 & 1.437 & 1.338 & 1.369 & 1.120 & 1.212 & 0.691 & 1.009 \\
$e_{kin}$
  & 0.788 & 0.774 & 0.815 & 0.802 & 0.926 & 0.872 & 1.240 & 0.974  \\
$e_{HO}$
  & 0.723 & 0.726 & 0.691 & 0.702 & 0.613 & 0.646 & 0.472 & 0.579 \\
$e_{\rho}$
  &-0.039 &-0.032 &-0.084 &-0.070 &-0.209 &-0.153 &-0.511 &-0.272 \\
\hline
\end{tabular}
\end{center}
% >>>>>>>>>>>>>>>>>>>>>>>>>>>>>>>>>>>>>>>>>>>>>>>>>>>>>>>>>>>>>>>>>>
\pagebreak
\section*{Figure Captions}

Figure 1. Density (nomalized to unity) of 200 atoms in a spherical trap
in $a_{HO}^{-3}$ units)
 as a function of the distance (in $a_{HO}$ units)
in the repulsive case calculated from the solution of the GPE (solid
line) and using the TF approach described in the text (dashed line).

Figure 2. The same as Figure 1 but with 200000 atoms in the trap.

Figure 3. Self-consisten potential (in $\hbar
\omega$ units) corresponding to a spherical trap containing 200 atoms
as a function of the distance ($a_{HO}$ units) in the repulsive case
calculated from the solution of the GPE (solid
line) and using the TF approach described in the text (dashed line).

Figure 4. The same as Figure 3 but with 200000 atoms in the trap.

Figure 5. The same as Figure 1 but with 250 atoms in the trap in the
attractive case.

Figure 6. The same as Figure 1 but with 1500 atoms in the trap in the
attractive case.

Figure 7. The same as Figure 3 but with 250 atoms in the trap in the
attractive case.

Figure 8. The same as Figure 3 but with 1500 atoms in the trap in the
attractive case.

Figure 9. Kinetic energy density per particle of a spherical trap (in
$\hbar \omega$ $a_{HO}^{-3}$ units)
containing 200000 atoms as a function of the distance (in $a_{HO}$
units)
in the repulsive case calculated from the solution of the GPE (solid
line) and using the TF approach described in the text (dashed line).

Figure 10. The same as Figure 9 but with 250 atoms in the trap in
the attractive case.

\pagebreak
\end{document}